\documentclass[hyper]{JHEP} % 10pt is ignored!

\usepackage{epsfig}

\newcommand\fverb{\setbox\pippobox=\hbox\bgroup\verb}
\newcommand\fverbdo{\egroup\medskip\noindent%
                        \fbox{\unhbox\pippobox}\ }
\newcommand\fverbit{\egroup\item[\fbox{\unhbox\pippobox}]}
\newbox\pippobox
\newcommand{\newc}{\newcommand}
\newc\eg{{\it {e.g.}}}  \newc\etal{{\it {et al.}}} \newc\ie{{\it i.e.}}
\newc\etc{{\it {etc}}}  

\newcommand\lsim{\mathrel{\rlap{\lower4pt\hbox{\hskip1pt$\sim$}}
    \raise1pt\hbox{$<$}}}
\newcommand\gsim{\mathrel{\rlap{\lower4pt\hbox{\hskip1pt$\sim$}}
    \raise1pt\hbox{$>$}}}

\newc{\mhalf}{m_{1/2}}      \newc{\mzero}{m_0}

\newc{\tanb}{\tan\beta}
\newc{\azero}{A_0}
\newc{\at}{A_t} \newc{\abot}{A_b} \newc{\atau}{A_\tau} 

\newc{\bmu}{B\mu}           \newc{\sgn}{{\rm sgn}}
\newc{\mone}{M_1}           \newc{\mtwo}{M_2}

\newc{\charone}{\chi_1^\pm} \newc{\mcharone}{m_{\chi_1^\pm}}

\newc{\hl}{h}               \newc{\mhl}{m_{\hl}}
\newc{\hh}{H}               \newc{\mhh}{m_{\hh}}
\newc{\ha}{A}               \newc{\mha}{m_{\ha}}
\newc{\hc}{H^{\pm}}         \newc{\mhc}{m_{\hc}}

\newc{\qzero}{Q_0}          \newc{\qstop}{Q_{\widetilde t}}

\newc{\amu}{a_{\mu}}        \newc{\amususy}{a_{\mu}^{\rm SUSY}}
\newc{\amuexpt}{a_{\mu}^{\rm expt}}        \newc{\amusm}{a_{\mu}^{\rm SM}}
\newc{\deltaamususy}{\Delta a_{\mu}^{\rm SUSY}}

\newc{\msbar}{\overline {\rm MS}} \newc{\drbar}{\overline {\rm DR}}

\newc{\mt}{m_t} \newc{\mb}{m_b} \newc{\mtau}{m_{\tau}}
\newc{\yt}{h_t} \newc{\yb}{h_b} \newc{\ytau}{h_{\tau}}
\newc{\mtpole}{m_t^{\rm pole}} \newc{\mbpole}{m_b^{\rm pole}} 
\newc{\mtaupole}{m_{\tau}^{\rm pole}} 

\newc{\mtmtsmmsbar}{m_t(m_t)^{\msbar}_{{\rm SM}}}
\newc{\mtmtsmdrbar}{m_t(m_t)^{\drbar}_{{\rm SM}}}
\newc{\mtmtmssmdrbar}{m_t(m_t)^{\drbar}_{{\rm SUSY}}}

\newc{\mbmbsmmsbar}{m_b(m_b)^{\msbar}_{{\rm SM}}}
\newc{\mbmzsmmsbar}{m_b(\mz)^{\msbar}_{{\rm SM}}}
\newc{\mbmzsmdrbar}{m_b(\mz)^{\drbar}_{{\rm SM}}}
\newc{\mbmzmssmdrbar}{m_b(\mz)^{\drbar}_{{\rm SUSY}}}
\newc{\mtaumzsmmsbar}{m_{\tau}(\mz)^{\msbar}_{{\rm SM}}}
\newc{\mtaumzsmdrbar}{m_{\tau}(\mz)^{\drbar}_{{\rm SM}}}
\newc{\mtaumzmssmdrbar}{m_{\tau}(\mz)^{\drbar}_{{\rm SUSY}}}

\newc{\mgut}{M_{\rm GUT}}

\newc{\mplanck}{M_{\rm P}}      \newc{\mpl}{M_{\rm Pl}}
\newc{\msusy}{M_{\rm SUSY}}      \newc{\ms}{M_{\rm S}}

\newc{\jxf}{J({\xf})}
\newc{\jxfexact}{J_{\rm exact}({\xf})}  \newc{\jxfexp}{J_{\rm exp}({\xf})}
\newc{\VEV}[1]{\langle #1 \rangle}

\newc{\xf}{x_f}
\newc\vrel{v_{\rm rel}}
\newcommand\mchi{m_{\chi}}              

\newc\sell{{\widetilde e}_L}      \newc\msell{m_{\sell}}
\newc\selr{{\widetilde e}_R}      \newc\mselr{m_{\selr}}

\newc\snue{{\widetilde \nu}_e}      \newc\msnue{m_{\snue}}
\newc\snutau{{\widetilde \nu}_\tau}      \newc\msnutau{m_{\snutau}}

\newc\supl{{\widetilde u}_L}      \newc\msupl{m_{\supl}}
\newc\supr{{\widetilde u}_R}      \newc\msupr{m_{\supr}}
\newc\sdl{{\widetilde d}_L}      \newc\msdl{m_{\sdl}}
\newc\sdr{{\widetilde d}_R}      \newc\msdr{m_{\sdr}}

\newcommand\stopq{{\widetilde t}}   \newcommand\mstopq{m_{\stopq}}

\newcommand\stopone{{\widetilde t}_1}   \newcommand\mstopone{m_{\stopone}}
\newcommand\stoptwo{{\widetilde t}_2}   \newcommand\mstoptwo{m_{\stoptwo}}
\newcommand\stopl{{\widetilde t}_L}   \newcommand\mstopl{m_{\stopl}}
\newcommand\stopr{{\widetilde t}_R}   \newcommand\mstopr{m_{\stopr}}

\newcommand\sbotone{{\widetilde b}_1}   \newcommand\msbotone{m_{\sbotone}}
\newcommand\sbottwo{{\widetilde b}_2}   \newcommand\msbottwo{m_{\sbottwo}}

\newcommand\stauone{{\widetilde \tau}_1}   \newcommand\mstauone{m_{\stauone}}
\newcommand\stautwo{{\widetilde \tau}_2}   \newcommand\mstautwo{m_{\stautwo}}

\newcommand\mgluino{m_{\widetilde g}}

\newc\hpm{H^\pm} \newc\hp{H^+} \newc\hm{H^-} 
\newc\sfermion{\tilde f}  \newc\msfermion{m_{\sfermion}}  

\newc\second{{\rm sec}} 

\newc\alphas{\alpha_s}

\newc\alphaem{\alpha_{em}}

\newc{\gstar}{g_\ast}           \newc{\gsstar}{g_{s\ast}}
\newc{\geff}{g_{\rm eff}}

\newcommand\mz{m_{Z}}

\newc{\sthw}{\sin\theta_W}              \newc{\cthw}{\cos\theta_W}
\newc{\bino}{\widetilde B}              \newc{\wino}{\widetilde W_3}
\newc{\higgsinob}{{\widetilde H}^0_b}   \newc{\higgsinot}{{\widetilde H}^0_t}

\newc{\abund}{\Omega h^2}
\newc{\abundchi}{\Omega_\chi h^2}
\newc{\abundcdm}{\Omega_{{\rm CDM}} h^2}
\newc{\omegam}{\Omega_{{\rm M}}}       \newc{\abundm}{\Omega_{{\rm M}} h^2}
\newc{\omegab}{\Omega_{{\rm b}}}	\newc{\abundb}{\Omega_{{\rm b}} h^2}
\newc{\omegacdm}{\Omega_{{\rm CDM}}}   \newc{\omegatot}{\Omega_{{\rm TOT}}}

\newc{\rhocrit}{\rho_{crit}}
\newc{\rhochi}{\rho_{\chi}}

\newcommand\tev{\,\mbox{TeV}}
\newcommand\gev{\,\mbox{GeV}}
\newcommand\mev{\,\mbox{MeV}}

\newc\br{\mbox{BR}}

\newc{\ra}{\rightarrow}
\newc{\beq}{\begin{equation}}
\newc{\eeq}{\end{equation}}
\newc{\bea}{\begin{eqnarray}}
\newc{\eea}{\end{eqnarray}}
\renewcommand\[{\left[}
\renewcommand\]{\right]}
\long\def\begincomment#1\endcomment{%
        \begingroup\sf\baselineskip12pt#1\endgroup}
\title{New Cosmological and Experimental Constraints on the CMSSM}

\author{Leszek Roszkowski\\
        TH Division, CERN, CH-1211 Geneva 23, Switzerland and\\
        Department of Physics, Lancaster University, 
        Lancaster LA1 4YB, England\\
        E-mail: \email{Leszek.Roszkowski@cern.ch}}

\author{Roberto Ruiz de Austri\\
        Department of Physics, Lancaster University, Lancaster LA1
4YB, England\\
        E-mail: \email{r.ruizdeaustri@lancaster.ac.uk}}

\author{Takeshi Nihei\\
        571-B, College of Science and Technology, Nihon University,
 1-8-14, Kanda-Surugadai, Chiyoda-ku, Tokyo, 101-8308, Japan\\
        E-mail: \email{nihei@phys.cst.nihon-u.ac.jp}}

\received{\today}              %%
\preprint{\hepph{0106334}}
\preprint{CERN--TH/2001-163}

\abstract{ We analyze the implications of several recent cosmological and
experimental measurements for the mass spectra of the Constrained
MSSM (CMSSM). We compute the relic abundance of the neutralino 
and compare the new
cosmologically expected and excluded mass ranges with those
ruled out by the final LEP bounds on the lightest chargino and Higgs
masses, with those excluded by current experimental values of
$\br( B\ra X_s \gamma )$, and with those
favored by the recent measurement of the anomalous magnetic moment of
the muon. We find that for $\tanb\lsim45$ there remains relatively
little room for the mass spectra to be consistent with the interplay of
the several constraints. On the other hand, at larger values of
$\tanb$ the decreasing mass of the pseudoscalar Higgs gives rise
to a wide resonance in the neutralino WIMP pair-annihilation, whose
position depends on the ratio of top and bottom quark masses. 
As a consequence, 
the cosmologically expected regions consistent with
other constraints often grow significantly and generally shift towards
superpartner masses in the~$\tev$ range. }


\begin{document} 

%%%%%%%%%%%%%%%%%%%%%%%%%%%%%%%%%%%%%%%%%%%%%%%%%%%%%%%%%%%%%%%%%%%%%%%%%%%%
\section{Introduction}\label{intro:sec}

Cosmology provides an important restriction on otherwise allowed
supersymmetric mass spectra. This in particular is the case with the relic
density of the lightest supersymmetric particle (LSP) which, in the
presence of $R$-parity, is stable. A natural candidate for the LSP is
the lightest neutralino $\chi$. Its relic abundance $\abundchi$ has
been a subject of a large volume of papers~\cite{jkg96:ref}, starting
from Refs.~\cite{goldberg93,ehnos} up to the recent comprehensive
studies~\cite{efgos01,eno01,ads0102,bk01}.  It is well known that $\abundchi$
can vary over several orders of magnitude but that it is also often
consistent with the abundance of non-baryonic cold dark matter (CDM)
in the Universe whose determinations have been improving steadily.
Recent reviews give more narrow ranges for the components (matter
$\omegam$ and baryonic $\omegab$) of the Universe and the Hubble
parameter $h$ than in the past.  For example, for the matter component
in Ref.~\cite{turnerjune01} one finds $\omegam=0.33\pm0.035$ which is
consistent with $0.4\pm0.1$ of Ref.~\cite{primack00} and
$0.35\pm0.05$ of Ref.~\cite{krauss01}. The baryonic component is now
$\abundb\simeq0.02$~\cite{turnerjune01,primack00,krauss01}
while the Hubble parameter is $h=0.72\pm0.07$~\cite{turnerjune01}
($0.65\pm0.08$~\cite{primack00}). Based on this, and assuming that the
neutralino LSP makes up the dominant component of the CDM in the
Universe, we now select the range of the neutralino relic abundance
\beq
0.1\lsim\abundchi\lsim0.2
\label{newachi:eq}
\eeq
as conservatively matching the recent observations. Furthermore, we
put an upper bound
\beq
\abundchi\lsim0.3
\label{newachiubound:eq}
\eeq 
based solely on the constraints on the age of the Universe. These ranges
will also allow a comparison with the favored values of $\abundchi$ in
the range $0.1\lsim\abundchi\lsim0.3$ which have often been used in
the recent literature. 

The latest determinations of $\omegacdm$ from the
measurements of the CMBR, based on assuming reasonable ranges for
other cosmological parameters (priors), like $\omegatot=1$, \etc, give
$\abundcdm\simeq
0.14\pm0.04$~\cite{prykeetal01:ref,netterfieldetal01:ref}, or, assuming
more priors, even much smaller errors: $\abundcdm\simeq
0.13\pm0.01$~\cite{netterfieldetal01:ref}. We do not feel yet ready to
accept these narrow ranges as robust enough for our analysis. In
particular, we note that, assuming $h=0.72\pm0.08$, the DASI
analysis~\cite{prykeetal01:ref} finds $\omegam=0.40\pm0.15$ which
implies $0.1<\abundm<0.35$. Likewise, the most recent ROSAT
measurement gives $\omegam=0.35\pm0.12$~\cite{rosat01} which again
allows for larger values of $\abundm$.
Nevertheless, below we will discuss the
impact on our results of assuming, after subtracting the baryonic
component, that $\abundchi$ is in the range $0.10<\abundchi<0.12$ and
$0.10<\abundchi<0.15$.

Cosmological constraints (\ref{newachi:eq})--(\ref{newachiubound:eq})
have a particularly strong effect in the framework of the Constrained
MSSM (CMSSM)~\cite{nillesrev,kkrw94:ref}. In the CMSSM, in addition to the
requirement of a common gaugino mass $\mhalf$ at the unification scale
$\mgut$, which is usually made in the more generic Minimal
Supersymmetric Standard Model (MSSM), one further assumes that the
soft masses of all scalars (sfermion and Higgs) are equal to $\mzero$
at $\mgut$, and analogously that the trilinear soft terms unify at
$\mgut$ at some common value $\azero$. These parameters are run using
their respective Renormalization Group Equations (RGEs) from $\mgut$
to some appropriately chosen low-energy scale $\qzero$ where the Higgs
potential (including full one-loop corrections) is minimized while
keeping the usual ratio $\tanb$ of the Higgs VEVs fixed. The
Higgs/higgsino mass parameter $\mu$ and the bilinear soft mass term $\bmu$
are next computed from the conditions of radiative electroweak
symmetry breaking (EWSB), and so are the Higgs and superpartner masses. The
CMSSM thus has a priori only the usual
\beq 
\tanb,\ \ \mhalf,\ \ \mzero,\ \ \azero,\ \ \sgn(\mu) 
\eeq 
as input parameters. (A comprehensive set of formulae that we will
refer to can be found for example in Ref.~\cite{tevatrontworep00}.)
However, in the case of large $\mhalf,\mzero\gsim1\tev$ and/or large
$\tanb\sim {\cal O}(\mt/\mb) $ some resulting masses will in general
be highly sensitive to the assumed physical masses of the top and the
bottom (as well as the tau)~\cite{efgos01} but they will {\em also
strongly depend on the correct choice} of the scale $\qzero$. This in
particular will affect the impact of the cosmological constraints
(\ref{newachi:eq})--(\ref{newachiubound:eq}) as we will discuss below.

In the CMSSM, the LSP neutralino is often a nearly pure
bino~\cite{na92,rr93,kkrw94:ref} because the requirement of radiative
EWSB typically gives $|\mu|\gg \mone$ where $\mone$ is the soft mass
of the bino.  This often (albeit not
always!~\cite{kkrw94:ref}) allows one to impose strong constraints from
$\abundchi< {\cal O}(1)$ on $\mhalf$ and $\mzero$ (and therefore also
on heaviest Higgs and superpartner masses) in the ballpark of $1\tev$.
This was originally shown in Refs.~\cite{rr93,kkrw94:ref} and later
confirmed by many subsequent studies starting from~\cite{baerbrhlik96}
up to the most recent analyses~\cite{efgos01,eno01,fmm99,bk01}.

%%%%%%%%%%%%%%%%%%%%%%%%%%%%%%%%%%%%%%%%%%%%%%%%%%%%%%%%%%%%%%%%%%%%%%%%%%
%
\section{Experimental Constraints}\label{expt:sec}

In the case of the CMSSM, the most important experimental
constraints from LEP are those on the masses of the lightest chargino
$\charone$ and Higgs boson $\hl$. For the first one we adopt the bound
$\mcharone>104\gev$ since the actual limits from the LEP experiments
are very close to this value. The lightest Higgs mass has been a
subject of much debate on both the experimental~\cite{lephiggsclaim00}
and the theory side. The
termination of LEP has left the first one unresolved for at least a
few years. On the theory front, due to large radiative corrections, the
precise value of $\mhl$ still remains somewhat dependent on the
procedure of computing it~\cite{cqw,espinosa,chhhww00,feynhiggsfast:ref}.
In our analysis we will conservatively
assume $\mhl>113\gev$ but will keep in mind that the theoretical
uncertainty in $\mhl$ in the CMSSM is probably of the order of $2$--$3\gev$.

Non-accelerator experimental results are also of much importance.
First, there has been much recent activity in determining $\br( B\ra
X_s \gamma )$. A recent combined experimental
result~\cite{misiakmoriond01:ref}\footnote{The most recent
update~\cite{gm01}, which incorporates the new CLEO
result~\cite{cleo0108}, gives $\br( B\ra X_s \gamma )
=(3.23\pm0.42)\times 10^{-4}$.  As we will coment again later, adopting
this new data would exclude somewhat larger regions of the CMSSM parameter
space.}  $\br( B\ra X_s \gamma ) = (3.11\pm 0.39)\times 10^{-4}$
allows for some, but not much, room for contributions from SUSY when
one compares it with the updated prediction for the Standard Model
(SM) $\br( B\ra X_s \gamma ) = (3.73\pm 0.30)\times
10^{-4}$~\cite{misiakmoriond01:ref}. Second, at large $\tanb$
next-to-leading order supersymmetric corrections to $b\ra s\gamma$
become important~\cite{hrs:ref,bb98,br99,bhgk00,cgnw00:ref,dgg00:ref}.
In our analysis we adopt the full expressions for the dominant terms
derived in Ref.~\cite{dgg00:ref}. We also include the $b$-quark mass
effect on the SM value which was subsequently pointed out in
Ref.~\cite{gm01}.  We add the two $1\sigma$ errors (the experimental
and SM) in quadrature and further add linearly $0.2$ to accommodate
the theoretical uncertainty in SUSY contributions which is roughly
$5\%$ of the SM value for branching ratio~\cite{gambinopc}. Altogether
we conservatively allow our results to be in the range $\br( B\ra X_s
\gamma ) = (3.11\pm 0.69)\times 10^{-4}$ for SM plus two-Higgs
doublets plus superpartner contribution. The excluded regions of SUSY
masses will not however be extremely sensitive to the choice of these
error bars.

Lastly, much excitement has recently been caused by the first
measurement by the Brookhaven experiment E821 of the anomalous
magnetic moment of the muon $\amu=(g_\mu-2)/2$~\cite{e821jan01:ref}.
Taken at face value, the result implies a $2.6\sigma$ discrepancy
between the experimental value and the SM prediction
$\amuexpt-\amusm=(43\pm16)\times10^{-10}$. There is much ongoing debate about
improving the understanding of the precise contribution from the SM,
as has been reported for example in
Ref.~\cite{czarneckimoriond01:ref}. In particular, there has been
a tendency of moving the SM prediction towards the range of
experimental values~\cite{narison01:ref}. In our analysis we will use
the published results to allow SUSY contributions in the ranges
$27\times10^{-10}<\deltaamususy<59\times10^{-10}$ ($1\sigma$) and
$11\times10^{-10}<\deltaamususy<75\times10^{-10}$ ($2\sigma$) but will
comment on the effect of varying these numbers later. Here we only
note that we consider the upper limit on $\deltaamususy$ (which implies a
{\em lower} limit on $\mhalf$ and $\mzero$) as rather robust. In
contrast, the lower limit on $\deltaamususy$ (and the resulting
{\em upper} limit on $\mhalf$ and $\mzero$) should be approached with
much caution. We will comment on this further when we present our
results.

%% In the rest of the paper we will first describe in more detail the
%% procedure for deriving mass spectra in the CMSSM. Next, we will comment on our
%% method of computing the neutralino relic abundance. Then we will
%% present our results.
%% 

%%%%%%%%%%%%%%%%%%%%%%%%%%%%%%%%%%%%%%%%%%%%%%%%%%%%%%%%%%%%%%%%%%%%%%%%%%
%
\section{Procedure}\label{procedure:sec}

We calculate superpartner and Higgs 
mass spectra using the package ISASUGRA~(v.7.51) but
make some important modifications which will be described below. We
refer the reader to Ref.~\cite{isasugra:ref} for a more detailed
description of the code. Here we will only highlight the main points
of the procedure. The overall strategy is to first find an approximate
spectrum of the Higgs and SUSY masses and then iterate the procedure of
running the RGEs between $\mgut$ and
the low scale until a satisfactory consistency is achieved. In the
initial step one runs the two-loop RGEs for the gauge
couplings up and identifies $\mgut$ as the point where $\alpha_1$ and
$\alpha_2$ are equal.
Exact unification of $\alphas$ with the other two SM gauge
couplings is not assumed as it would predict a value far
too large compared with the experimental range $\alphas(\mz)=0.1185
\pm 0.002$~\cite{pdb01:ref} which we take as input. 

The top, bottom and tau masses must also be treated
with care since their assumed values often have an important effect on
the running of the RGEs, especially at large $\tanb$. The pole mass of
the top $\mtpole=174.3\pm5.1$~\cite{pdb01:ref} is initially
converted to the running mass $\mtmtsmmsbar$ in the $\msbar$ scheme using
a two-loop QCD correction with $\alphas$ running computed at three
loops, and is then identified with $\mtmtsmdrbar$ in the $\drbar$ scheme.
In subsequent
iterations, once SUSY masses have been computed, also one-loop SUSY
corrections from squark and gluino contributions are included 
to give $\mtmtmssmdrbar$.  

As regards the mass of the bottom, the Particle Data
Book~\cite{pdb01:ref} quotes the range $4.0$--$4.4\gev$ for the
running mass in the $\msbar$ scheme
$\mbmbsmmsbar$. We adopt the similar range
$\mbmbsmmsbar=4.25\pm0.15$~\cite{sachrajda00:ref} which has been used
in other recent analyses~\cite{efgos01,eno01}. 
In ISASUGRA $\mbmbsmmsbar$ is initially run
up to $\mz$ using a one-loop QCD+QED formula, with $\alphas$ running
computed at
three loops and $\alphaem$ at one loop, to obtain $\mbmzsmmsbar$ which
is assumed to be the same as in the $\drbar$-scheme. 
In subsequent steps, once
the SUSY masses have been computed, one includes SUSY
corrections~\cite{hrs:ref,pbmz97:ref} from squark/gluino and
squark/chargino loops which become important at large $\tanb$
\beq
\mbmzmssmdrbar= \mbmzsmdrbar \[ 1-\left(\frac{\Delta
    \mb}{\mb}\right)^{\rm SUSY}\],
\label{mbmssm:eq}
\eeq
where $\mbmzsmdrbar$ is computed in the initial step and
$\left({\Delta\mb}/{\mb}\right)^{\rm SUSY}$  stands for a SUSY mass correction.
(We note that in ISASUGRA it is the pole mass $\mbpole$ that is used as an
independent parameter and is actually `hardwired' at $\mbpole=4.9\gev$
[and in some routines at $\mbpole=5.0\gev$]. We have modified the
package to allow it to treat $\mbpole$ as an external parameter which
is then adjusted to match the desired value for $\mbmbsmmsbar$. The
conversion between the two is done using a one-loop expression.)

Finally, the
pole mass of the $\tau$-lepton is now well measured
$\mtaupole=1777.0\pm0.3\mev$~\cite{pdb01:ref}. It is converted to the
$\drbar$ scheme in SUSY using an analogous procedure to the bottom
mass. In particular, a threshold due to $\mb$ is added and one-loop
SUSY corrections due to just chargino exchange are included after the
initial step.

At $\mz$ the initial values of $\mbmzmssmdrbar$ and $\mtaumzmssmdrbar$
are used to compute the corresponding Yukawa couplings $\yb$ and
$\ytau$ for a fixed value of $\tanb$. The Yukawa coupling $\yt$ of the
top is computed at $\mt(\mt)$.  These couplings are next run up to $\mgut$
using two-loop RGEs in the $\drbar$ scheme with the same initial mass
threshold as for the gauge couplings, but are not assumed to unify.

In the second step, starting from $\mgut$, for a given choice of
$\tanb$, $\mhalf$, $\mzero$, $\azero$ and $\sgn(\mu)$, the RGEs for
the masses and couplings are run down to the low scale. Here much
attention is paid to extracting the parameters at a right energy scale
$\qzero$. The RGEs for the gaugino masses are run down until $Q$
reaches their respective running values and analogously for the
sfermion soft masses and the third-generation trilinear parameters
$A_t$, $A_b$ and $A_\tau$. The top Yukawa RGE is run down to
$\qzero=\mtpole$ 
while the RGEs for the other two Yukawa couplings of
the third generation are run down to $\mz$.

Of particular importance is a correct treatment of the Higgs sector and
the conditions for the EWSB. This is because of the spurious but
nagging $Q$-scale dependence of the $\msbar$ scheme~\cite{grz:ref}. Even
including full one-loop corrections to the Higgs potential is not
sufficient to significantly reduce the scale dependence when one minimizes the
Higgs potential at the scale $\mz$. Instead, it has been
argued, for example in~\cite{decarloscasas93:ref}, that the scale
dependence is significantly reduced by evaluating the (one-loop
corrected) Higgs potential at $\qstop\sim\sqrt{\mstopone \mstoptwo}$
with $\mstopone$ ($\mstoptwo$) denoting the physical masses of the stops.
This is because, at this scale, the role of the otherwise large
log-terms $\sim\log\left(\mstopq^2/Q^2\right)$ from the dominant
stop-loops will be reduced.
Here we follow this choice which is also adopted in
ISASUGRA. (Actually, in ISASUGRA a
very similar prescription is used with $\qstop=\sqrt{\mstopl \mstopr}$,
where $\mstopl$ ($\mstopr$) denote the soft masses of the stops, but
we do not think that this difference is of much importance). At
this scale one evaluates the conditions for the EWSB which determine $\mu^2$ 
as well as the bilinear soft mass parameter $\bmu$
\bea
\label{ewsbone:eq}
\mu^2&=&\frac{
\left(m_{H_d}^2+\Sigma_d^{(1)}\right) -
  \left(m_{H_u}^2+\Sigma_u^{(1)}\right)\tan^2\beta
}{\tan^2\beta-1} - \frac{1}{2}\mz^2\\ 
& & \nonumber \\
\label{ewsbtwo:eq}
2 B\mu&=& \tan 2\beta\left[ \left(m_{H_d}^2+\Sigma_d^{(1)}\right) -
  \left(m_{H_u}^2+\Sigma_u^{(1)}\right) \right] + \mz^2\sin 2\beta,
\eea
where $m^2_{H_{d,u}}$ are the squares of the 
soft mass terms of the Higgs doublets
$H_{d,u}$ and $\Sigma_{d,u}^{(1)}$ are their respective one-loop
corrections. 

In the package ISASUGRA only the (dominant) third-generation
(s)fermion one-loop corrections
were included. We have added {\em all} the one-loop corrections
following Ref.~\cite{pbmz97:ref}, in particular the 
chargino and neutralino ones. It has
been claimed in Ref.~\cite{klns00} that, while normally subdominant,
these can contribute to increasing $\mha$ at large $\tanb$ and large
$\mhalf$.

The mass of the pseudoscalar will play a crucial role in computing
$\abundchi$, especially at very large $\tanb\sim50$.
This is so for three reasons: 
$\mha$ becomes now much smaller~\cite{dn93} than at
smaller $\tanb$ due to the increased role of the bottom Yukawa coupling;
because the $A$-resonance in $\chi\chi\ra f\bar{f}$ is
dominant since the coupling $Af\bar{f}\sim \tanb$ for down-type
fermions; and because, in
contrast to the heavy scalar $\hh$, this channel is not $p$-wave
suppressed~\cite{jkg96:ref}. In ISASUGRA $\mha$ is computed as
\beq
\mha^2= \left(\tanb + \cot\beta\right)\left( -\bmu + \Delta_A^2\right)
\label{mhamass:eq}
\eeq
where $\Delta_A^2$ stands for 
the full one-loop corrections which can be
significant~\cite{pbmz97:ref}. 

The neutralino relic density can be reliably computed both away from
resonances and new final-state thresholds, where the usual expansion
in powers of $x=T/\mchi$ (where $T$ is the temperature) works well,
as well as in the vicinity of such special points. Exact analytic cross
sections, which will soon become available~\cite{nrr2}, allow us to
consistently and precisely determine $\abundchi$ both near and away
from such special points by solving the Boltzmann equation as in
Ref.~\cite{gg91,darksusy00}. (In the case of the expansion, analytic
formulae for the thermally-averaged product of the annihilation cross
section and the relative velocity are given in
Refs.~\cite{dn93,jkg96:ref} and are applicable far away from
resonances and thresholds. However, caution is advised in combining
them with a numerical integration of exact cross sections in the
vicinity of the resonances. This is because one then usually neglects
interference terms which can play some role.  Furthermore, the range
of $\mchi$ around the pseudoscalar resonance where the expansion fails
badly can be as large as several tens of $\gev$~\cite{nrr1}.)

We also include the co-annihilation with next-to-lightest SUSY
particles (NLSPs) which in some cases is important.  The
co-annihilation with the lightest chargino and the next-to-lightest
neutralino is treated without any approximation following
Refs.~\cite{eg97,darksusy00}. In the CMSSM, the lighter stau
($\stauone$) is the LSP in the region
$\mhalf\gg\mzero$~\cite{kkrw94:ref} and just above the boundary the
neutralino LSP can efficiently co-annihilate with it and with the
other sleptons~\cite{efos:coann}.  In this analysis we use the
approximate expressions given in the second paper of
Ref.~\cite{efos:coann} even though they do not include the effects of
$\ytau$, of the $\stauone-\stautwo$ mixing, in some channels of the
mass of the $\tau$, \etc~\cite{efgos01}, which make them less reliable
at large $\tanb\gsim20$~\cite{efgos01}. We have made an attempt at
improving the available formulae by including in the propagators the
widths of the gauge and Higgs bosons and the neutralinos, which
otherwise become singular. Furthermore, ISASUGRA does not include
one-loop corrections to the neutralino, chargino and slepton masses
which can be of the order of a few per cent~\cite{pbmz97:ref}. This
will have a comparable effect on the exact position of the boundary
between the neutralino and the $\stauone$ LSP especially at large
$\mhalf$. For these reasons we will treat our numerical results in the
co-annihilation region as somewhat less reliable but would not expect
any major changes if the mentioned effects were included.
\EPSFIGURE[t]{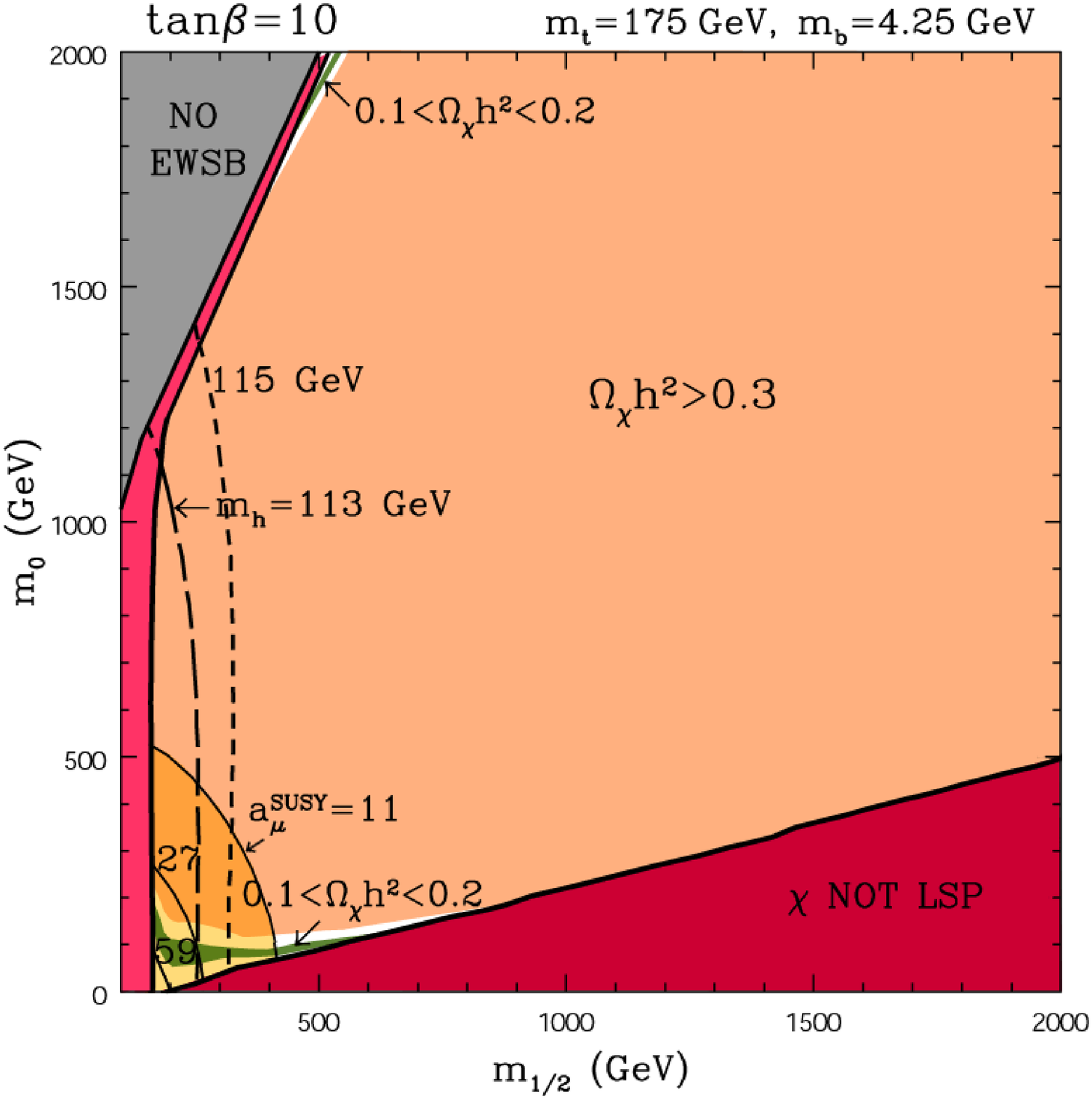,width=6in} 
{ The plane
($\protect{\mhalf},\protect{\mzero}$) for $\protect{\tanb=10}$,
$\azero=0$, $\mu>0$ and for $\mt\equiv\mtpole=175\gev$ and
$\mb\equiv\mbmbsmmsbar=4.25\gev$. The light red bands on the left are
excluded by chargino searches at LEP. In the grey wedge in the
left-hand corner electroweak symmetry breaking conditions are not
satisfied. The dark red region denoted `$\chi$ NOT LSP' corresponds to
the lighter stau being the LSP. The large light orange regions of
$\abundchi>0.3$ are excluded by cosmology while the narrow green bands
correspond to the expected range $0.1<\abundchi<0.2$. Also shown are
the semi-oval contours of $\amususy\equiv\deltaamususy/10^{-10}$
favored by the anomalous magnetic moment of the muon measurement at
$2\sigma$~CL ($\amususy=11,75$) and $1\sigma$~CL
($\amususy=27,59$). The $2\sigma$ range is shown in dark yellow. The
three lines in the figure correspond respectively to $\amususy=11,27$
and $59$, when moving towards the origin. The lines of the lightest
Higgs scalar mass $\mhl=113\gev$ and $115\gev$ are denoted by short
and long-dash lines, respectively.
\label{fig:tb10}
}

\EPSFIGURE[t]{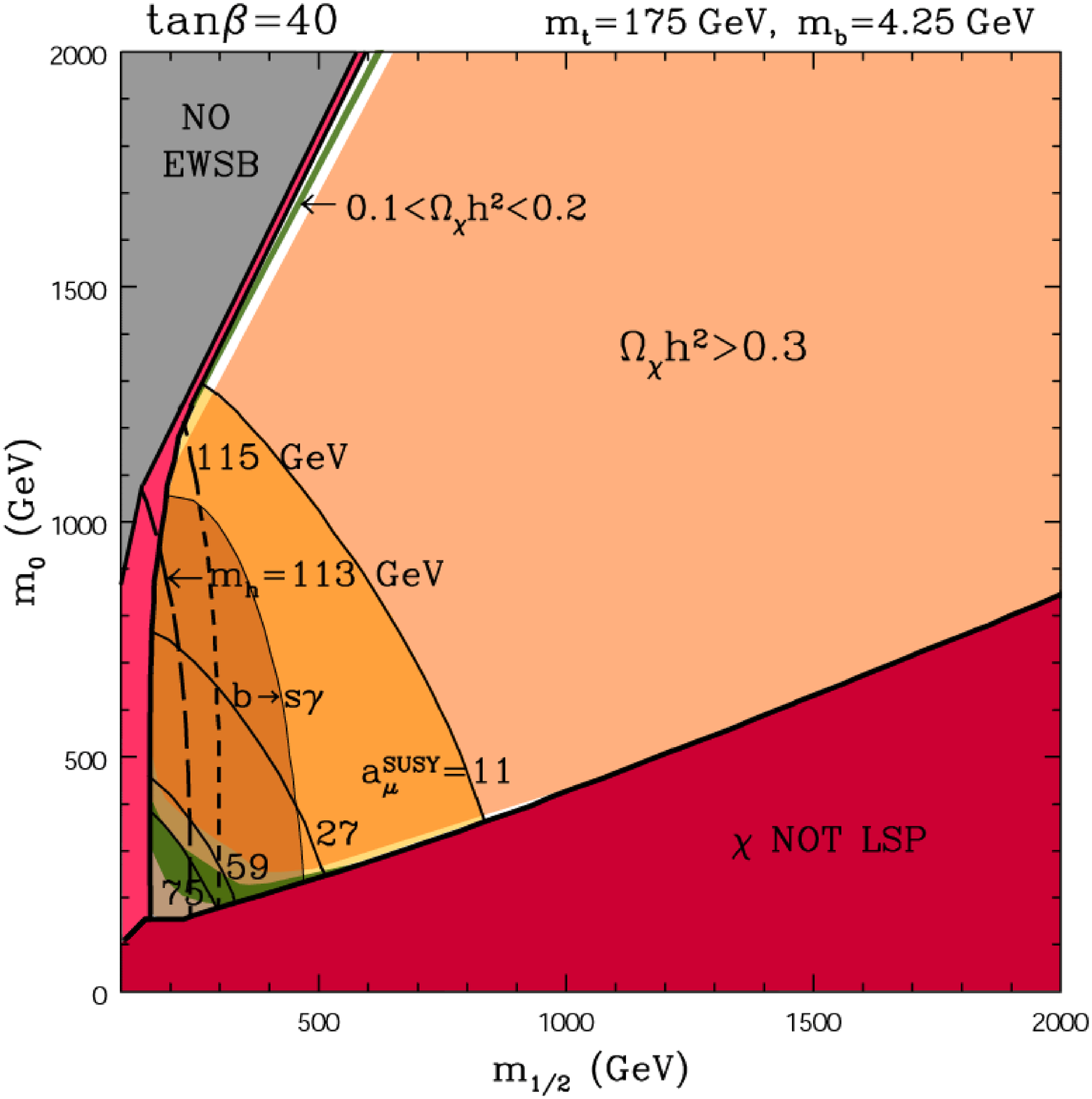,width=6in}
{
The same as in Fig.~\protect{\ref{fig:tb10}} but for $\tanb=40$. In
  addition, the light brown region now appears which is excluded by the
  lower bound of $\br(
  B\ra X_s \gamma ) = (3.11\pm 0.69)\times 10^{-4}$.
\label{fig:tb40}
}

\EPSFIGURE[t]{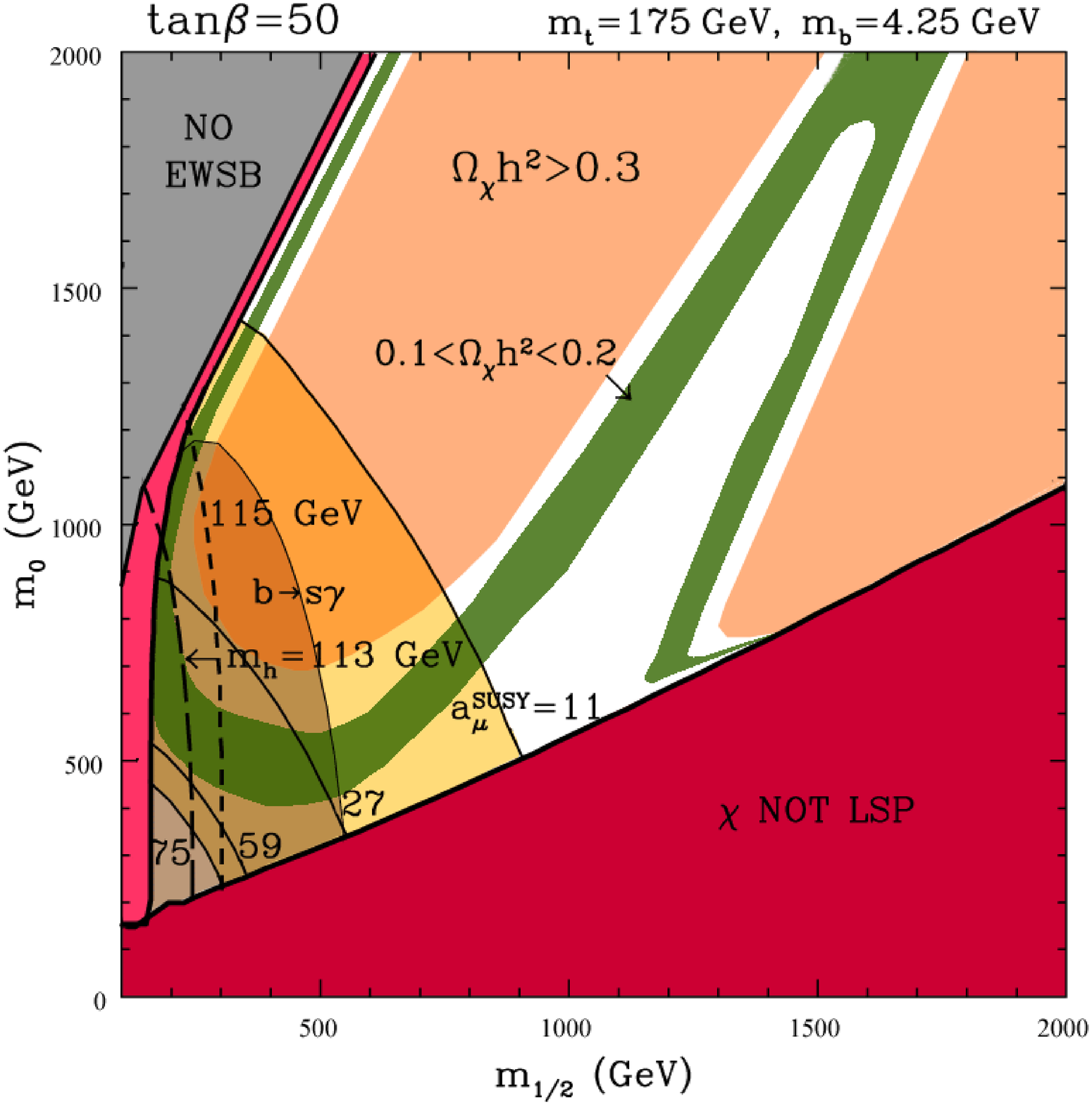,width=6in}
{
The same as in Fig.~\protect{\ref{fig:tb40}} but for
$\tanb=50$. The green band of the expected range
  $0.1<\abundchi<0.2$ has now changed considerably due to the
  appearance of a wide resonance $\chi\chi\ra \ha \ra f\bar{f}$. The
  white areas closer (further away) from the axes correspond to
  $\abundchi<0.1$ ($0.2<\abundchi<0.3$). 
\label{fig:tb50}
}

%%%%%%%%%%%%%%%%%%%%%%%%%%%%%%%%%%%%%%%%%%%%%%%%%%%%%%%%%%%%%%%%%%%%%%%%%%
%
\section{Results}\label{results:sec}

We present our results in the plane ($\mhalf,\mzero$) for several
representative choices of $\tanb$ and other relevant parameters as
specified below. First, in Figs.~\ref{fig:tb10}--\ref{fig:tb50} we
show the experimental and cosmological bounds for respectively
$\tanb=10,40$ and $50$ and for $\azero=0$, $\mu>0$ and the central
values of $\mt\equiv\mtpole=175\gev$ and
$\mb\equiv\mbmbsmmsbar=4.25\gev$. In Figs.~\ref{fig:tb10}
and~\ref{fig:tb40} we can see many familiar features. At
$\mhalf\gg\mzero$ there is a (dark red) wedge where the $\stauone$ is
the LSP. On the other side, at $\mzero\gg\mhalf$ we find large (grey)
regions where the EWSB is not achieved.  Just below the region of no-EWSB 
the parameter $\mu^2$ is small but positive which allows one to
exclude a further (light red) band by imposing the LEP chargino mass
bound. As one moves away from the wedge of no-EWSB, $\mu^2$ increases
rapidly. That implies that, just below the boundary of the no-EWSB
region, the LSP neutralino is higgsino-like but, as one moves away
from it, it very quickly becomes the usual nearly pure bino. This
causes the relic abundance $\abundchi$ to accordingly increase rapidly
from very small values typical for higgsinos in the hundred~$\gev$
range, through the narrow strip ($\Delta\mhalf\sim20\gev$) of the
cosmologically expected (green) range~(\ref{newachi:eq}) to much
larger values, excluded (light orange) by~(\ref{newachiubound:eq}). In
particular, in the whole region allowed by the chargino mass bound the
LSP is mostly bino-like.  We remind the reader that in the nearly pure
bino case the neutralino mass is given by $\mchi\simeq 0.44\mhalf
-2.8\sin 2\beta$~\cite{bkao98}.

It is clear from Figs.~\ref{fig:tb10} and~\ref{fig:tb40} that up until
$\tanb\sim40$ the overall shape of the cosmologically
expected~(\ref{newachi:eq}) and excluded~(\ref{newachiubound:eq})
regions does not change much. Generally one finds a robust (green)
region of expected $\abundchi$ at $\mhalf\sim\mzero$ in the range of a
few hundred~$\gev$~\cite{kkrw94:ref}. In addition, at
$\mhalf\gg\mzero$, just above the wedge where the LSP is the
$\stauone$, the co-annihilation of the neutralino LSP with $\stauone$
opens up a very long and very narrow strip which is allowed by the
bounds (\ref{newachi:eq})--(\ref{newachiubound:eq}).  Finally, as
mentioned above, at $\mzero\gg\mhalf$, very close to the region of
no-EWSB, again one finds a very narrow range of $\abundchi$ consistent
with (\ref{newachi:eq}). A closer examination would be required to
establish to what extent this region of $\mhalf$ and $\mzero$ overlaps
with that corresponding to `focus points' advocated in
Ref.~\cite{fmm99}.

\begin{figure}
\vspace*{-0.75in}
\hspace*{-.70in}
%\begin{center}
\begin{minipage}{8in}
\epsfig{file=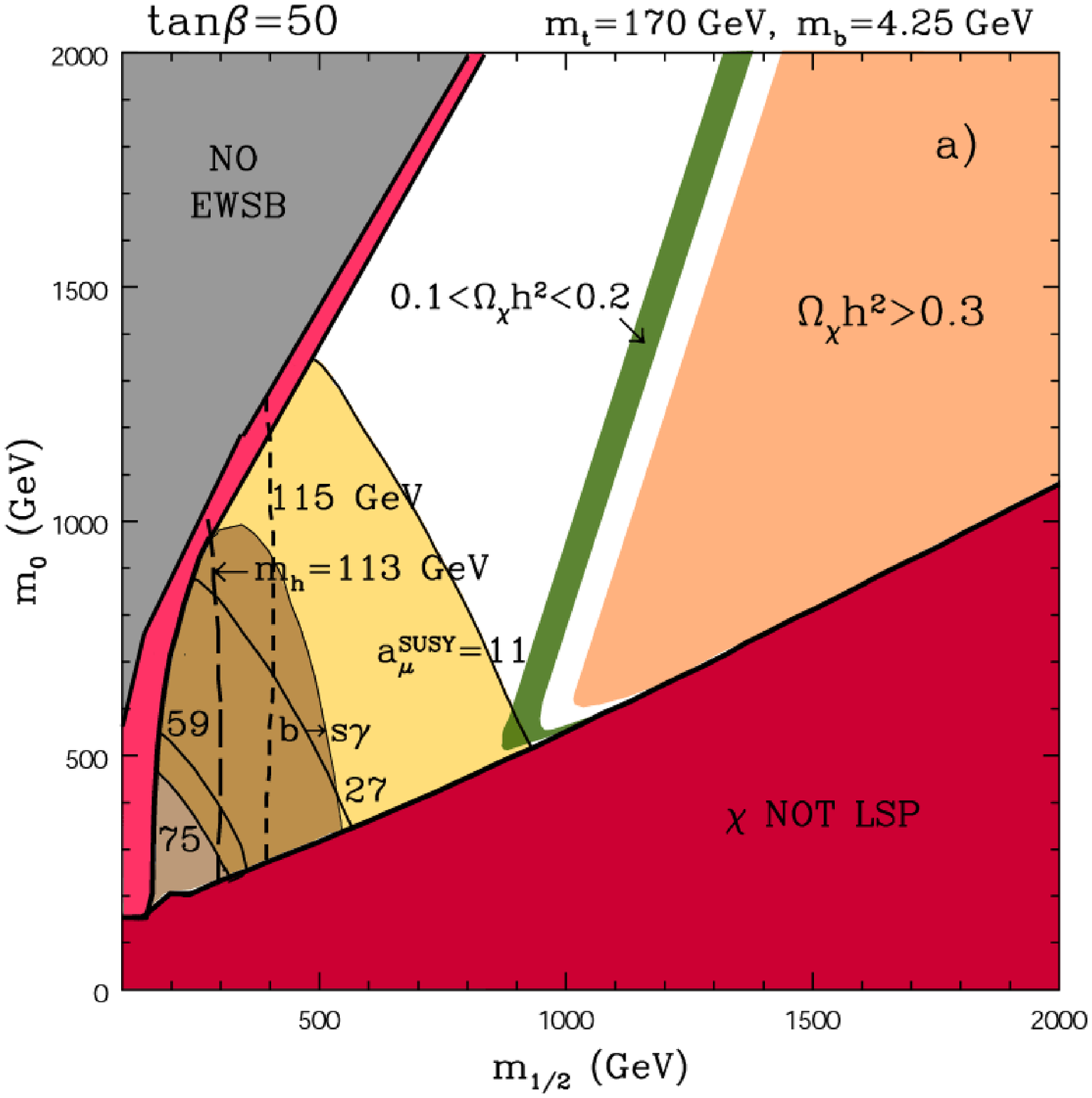,height=3.5in} %width=3.5in}
\hspace*{-0.18in}
\epsfig{file=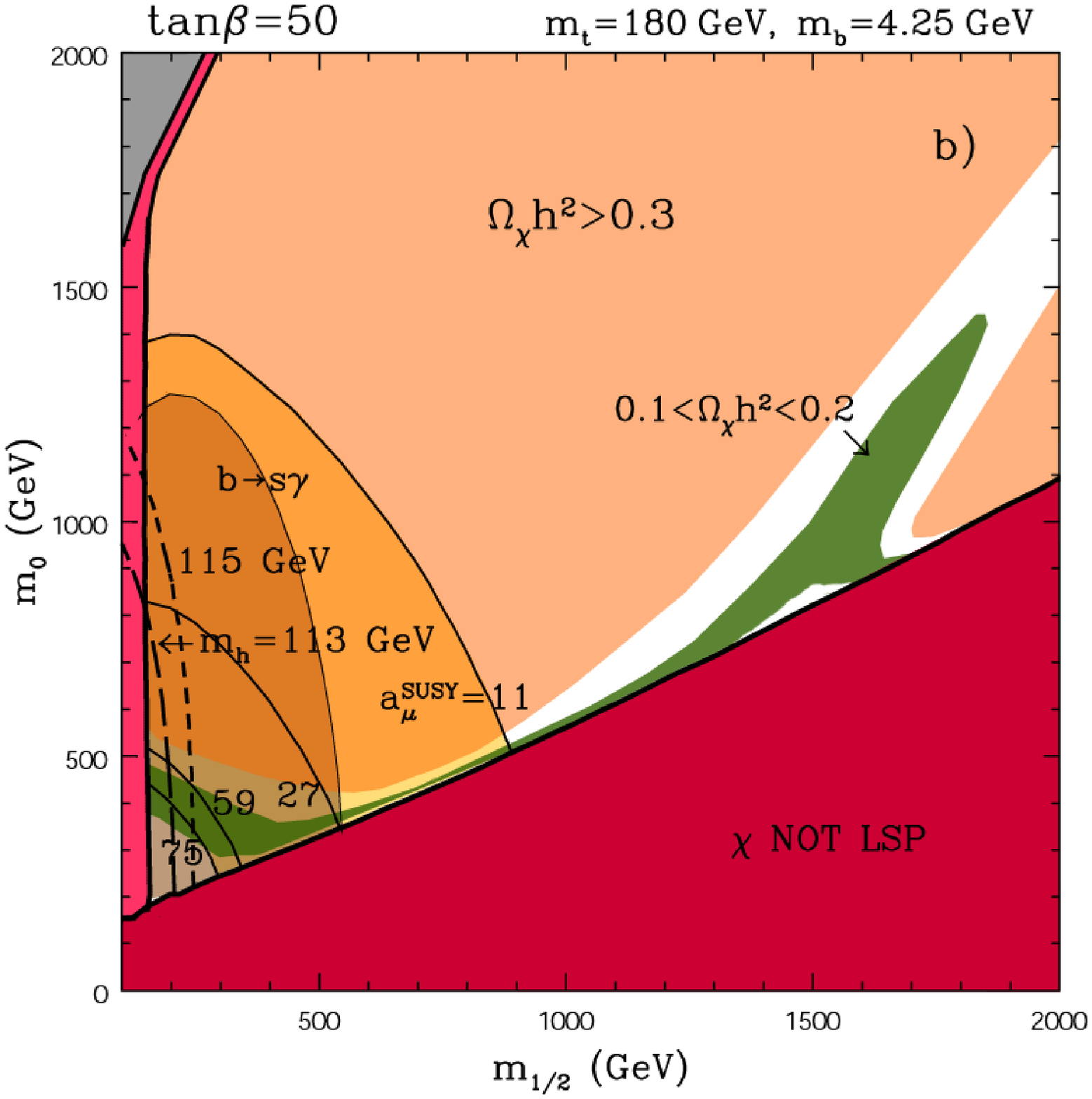,height=3.5in} %width=3.5in}
\end{minipage}
%\end{center}
\vspace*{.70in}
\hspace*{-.70in}
%\begin{center}
\begin{minipage}{8in}
\epsfig{file=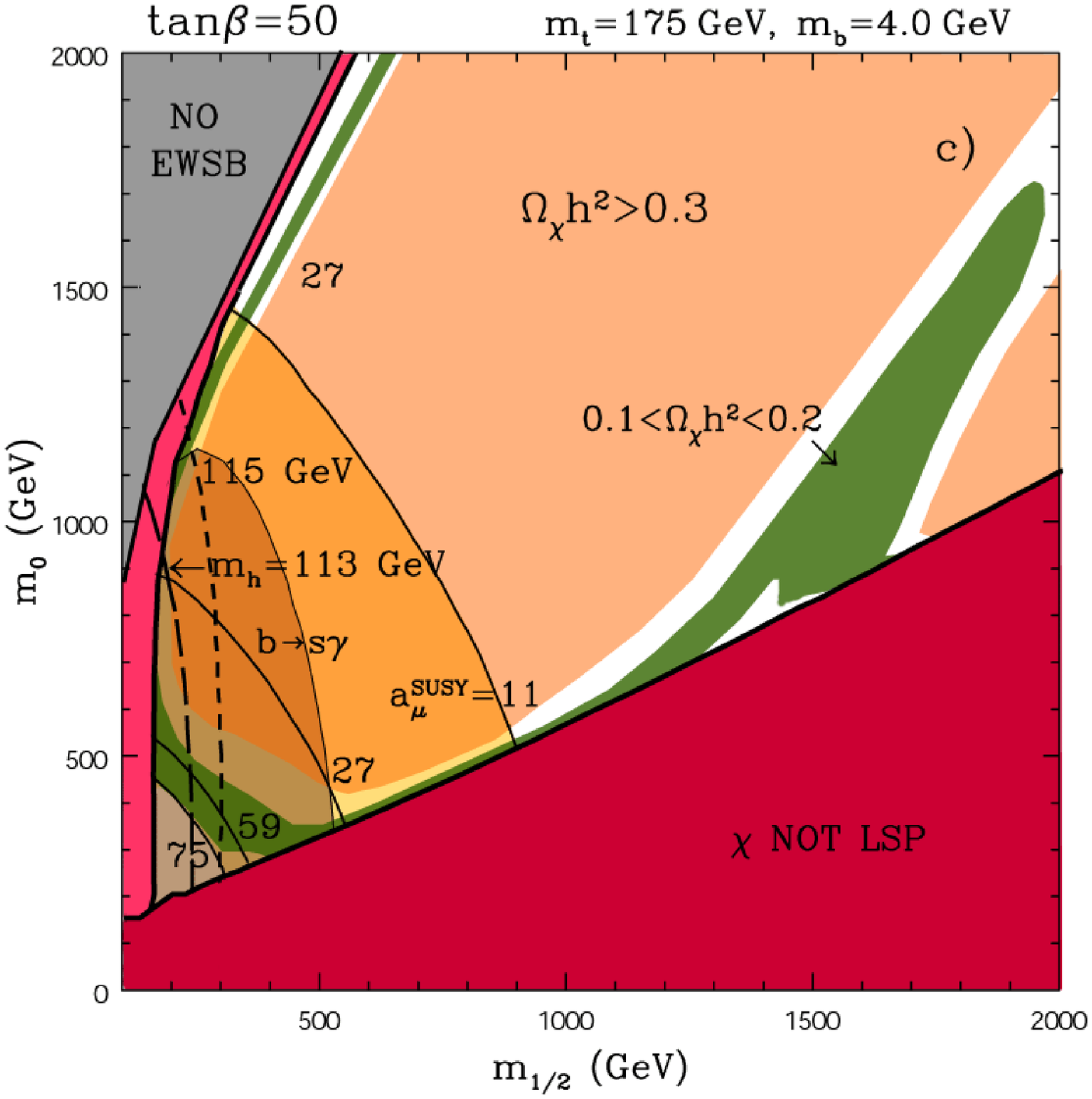,height=3.5in} %width=3.5in}
\hspace*{-0.18in}
\epsfig{file=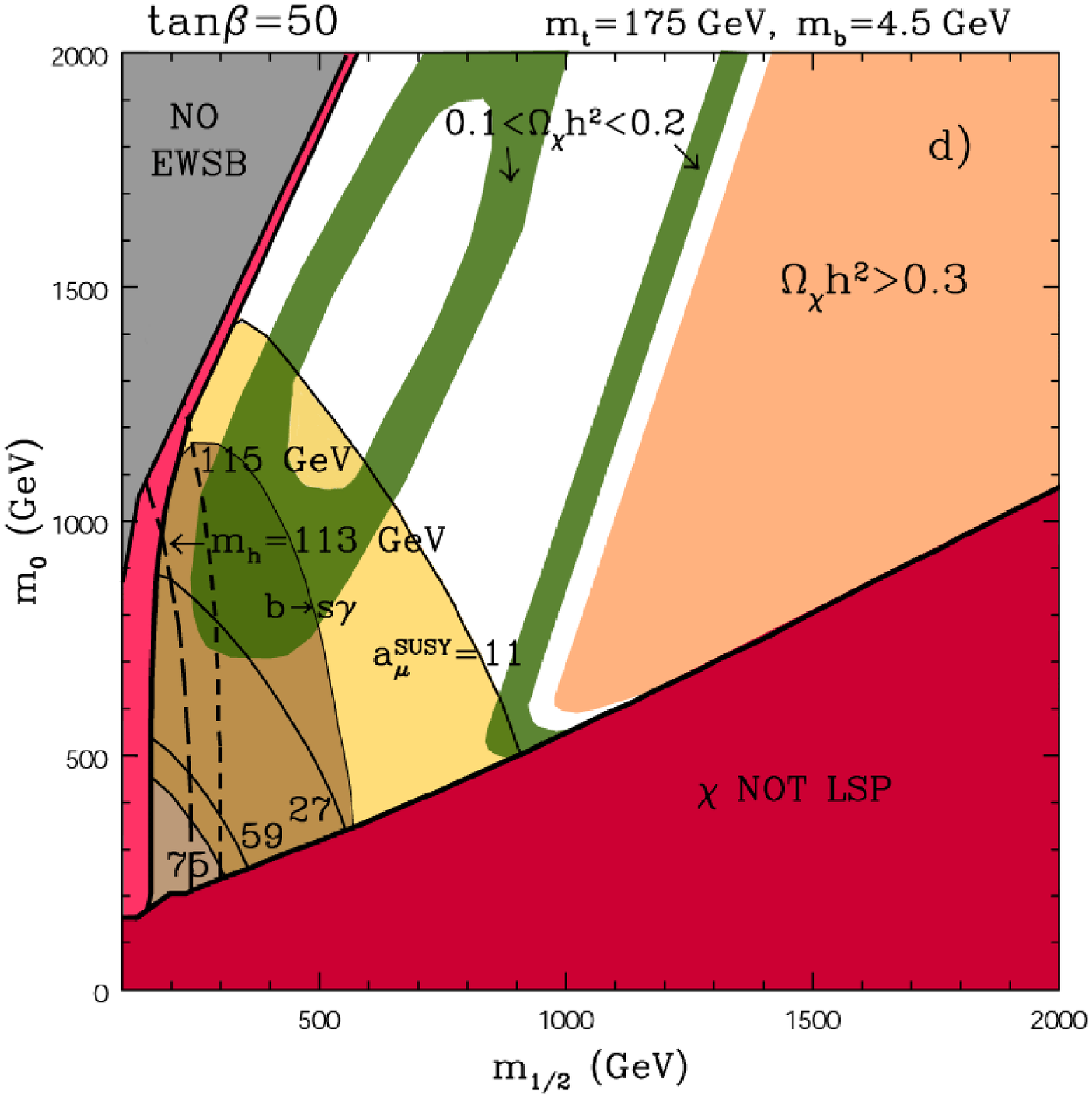,height=3.5in}
\end{minipage}
\caption{\label{fig:tb50detail} The same as in
Fig.~\protect{\ref{fig:tb50}} but for different values of the top and
bottom masses as indicated in respective windows.  In window~d) the
white area around the green `island' of the expected range
$0.1<\abundchi<0.2$ corresponds to $\abundchi<0.1$. The white strips
between the narrow green bands and the excluded (light orange) regions
of $\abundchi>0.3$ correspond to $0.2<\abundchi<0.3$, and so does
the white `hole' inside the green `island' in window~d).  }
%\end{center}
\end{figure}

The region of no-EWSB is quite sensitive to the relative values of
the top and bottom masses. Generally, at fixed $\tanb$, increasing
(decreasing) the top mass relative to the bottom mass causes the
region of no-EWSB to move up (down) considerably because of the
diminishing (growing) effect of the bottom Yukawa coupling on the loop
correction to the conditions of EWSB~(\ref{ewsbone:eq})--(\ref{ewsbtwo:eq}). 
This effect can be
explicitly seen in Fig.~\ref{fig:tb50detail}a and~b but it
remains basically true also for much smaller $\tanb$. At fixed top and
bottom masses, as $\tanb$ decreases, the region of no-EWSB moves
towards somewhat larger values of $\mzero$ and smaller values of
$\mhalf$ but the overall effect is not very significant. 

The lightest Higgs mass also grows with $\mt$ as
expected. In the Figures we plot the contours of $\mhl=113\gev$
(which, within the uncertainties mentioned earlier, 
corresponds to the LEP lower limit) and of $\mhl=115\gev$ where an
intriguing possibility of a Higgs signal has been
reported~\cite{lephiggsclaim00}. Larger values of $\mhl$ are given by
contours which are
shifted along the $\mhalf$ axis with roughly equal spacings but
diverge somewhat at larger $\mzero$. We remind
the reader that the values of $\mhl$ that we plot have been obtained
using ISASUGRA. When we use FeynHiggsFast~\cite{feynhiggsfast:ref}, we
obtain values lower by $\sim2-3\gev$, as can be seen from
Table~\ref{tableone}.

Other constraints behave with increasing $\tanb$ as expected. In
particular, the (light brown) region excluded by $\br( B\ra X_s \gamma
)$ grows significantly because the dominant chargino-squark contribution to
the branching ratio grows linearly with $\tanb$.  On the other hand,
the excluded region does not change much as we vary the assumed lower
bound on $\br( B\ra X_s \gamma )$. Taking the (slightly higher) 
most recent range (see Section~\ref{expt:sec}) enlarges the excluded
regions of the ($\mhalf,\mzero$)--plane by some additional $~50\gev$.

As for the SUSY contribution to the anomalous magnetic moment of the
muon, the $1\sigma$, and especially $2\sigma$, (yellow) range of
allowed $\deltaamususy$ becomes significantly enlarged towards larger
$\mhalf$ and $\mzero$ because the dominant sneutrino-chargino loop
exchange contribution grows with $\tanb$~\cite{chatnath96}. It is
clear that the dependence of $\deltaamususy$ on $\mhalf$ and $\mzero$ is
relatively weak and therefore, in our opinion, {\em no robust upper}
bound on superpartner masses can at present be placed. On the other
hand, the low ranges of $\mhalf$ and $\mzero$ below the lines
corresponding to the upper
$1\sigma$ ($2\sigma$) limits on $\deltaamususy=59,75\times10^{-10}$,
respectively, are probably firmly excluded but are not too big.

The region of very large $\tanb\sim 40 - 50$ deserves a separate
discussion. First, as can be seen by comparing Figs.~\ref{fig:tb10}
and~\ref{fig:tb40}, we find relatively little variation in the overall
shape of the constraints as $\tanb$ increases up to $\sim40$. 
However, as $\tanb$ grows, above $\sim 45$ a new important feature
appears which is prominently reflected in the neutralino relic
abundance, as can be seen from Figs.~\ref{fig:tb50}
and~\ref{fig:tb50detail}. The heavy Higgs bosons become `light'
enough~\cite{dn93,baerbrhlik96,bkao98,lns00} to ``enter the
($\mhalf,\mzero$)--plane from the right'' and cause a significant
decrease in $\abundchi$, predominantly through a very wide $\ha$-resonance.
Clearly, unlike for smaller values of $\tanb\lsim 45$, there is now much
variation in the position of the $\ha$-resonance and accordingly in
the shape of the allowed and excluded regions of $\abundchi$ as one
varies $\mt$ and/or $\mb$.  Generally, as $\mb$ increases
relative to $\mt$, the position of the resonance moves to the left
and the shapes of the regions where $\abundchi$ is 
in the prefered range~(\ref{newachi:eq}) and is excluded
by~(\ref{newachiubound:eq}) change accordingly.  
Qualitatively this behavior again clearly
demonstrates the growing impact of $\yb$ at large $\tanb$. More
quantitatively, one can see why this happens by examining 
Eq.~(\ref{mhamass:eq}) where the one-loop correction
$\Delta_A^2$~\cite{tevatrontworep00} to $\mha^2$  
from the sbottoms grows with $\tanb$ as $\mb^2/\cos^2\beta\sim\yb^2$. 

The white regions to the left of the green regions of
$0.1\lsim\abundchi\lsim0.2$, as well as around the green `island' in
Fig.~\ref{fig:tb50detail}d, correspond to 
$\abundchi<0.1$. The white areas
between the favored green regions and the excluded light orange range
($\abundchi>0.3$) as well as the white pocket inside the green region in
Fig.~\ref{fig:tb50detail}d correspond to $0.2<\abundchi<0.3$. One
should also mention that, in contrast to much smaller $\tanb$, at large
$\tanb\sim50$, $\abundchi$ does not grow rapidly with increasing
superpartner masses but instead changes very gradually and does not
exceed unity by more than a factor of a few. This is again
mostly caused by the very wide $\ha$-resonance because the $\ha$-width
is large and actually growing with $\tanb$ as well as $\mhalf$
and $\mzero$. As one cuts along the $\mhalf$ axis at fixed $\mzero$
in Figs.~\ref{fig:tb50} and~\ref{fig:tb50detail}, one can see that
$\abundchi$ first grows somewhat, then gradually decreases around the
$\ha$-resonance and then increases again at large $\mhalf$. 

The most recent measurements of $\omegam$ and $\omegacdm$ have now
implied much more restrictive ranges $0.10<\abundchi<0.12$ and
$0.10<\abundchi<0.15$, as discussed earlier. We do not show these
ranges in our Figures but can easily summarize their effect. The range
$0.10<\abundchi<0.15$ corresponds roughly to the half of the allowed
green strips (when cutting {\em along} them) on the side of the axes and the
origin. In Fig.~\ref{fig:tb50detail}d this corresponds to the the
outer halfs of the green `island'. Requiring $0.10<\abundchi<0.12$
causes a further shrinking of the green regions in the same direction
by roughly another factor of two.

Clearly, as $\tanb$ increases above $\sim45$, the cosmologically
expected and excluded regions are gradually shifted towards larger
$\mhalf$ and $\mzero$ thus significantly relaxing the tight bounds
characteristic of smaller $\tanb$. This is especially true when the
top-to-bottom mass ratio is on the lower side. Overall, however,
cosmological constraints on $\abundchi$ now permit much larger
superpartner masses than at smaller $\tanb$, and not only in the very
narrow strips close to the regions of no-EWSB and/or $\stauone$-LSP.

The case of very large $\tanb\gsim50$ is obviously quite complicated.
We are not even convinced that ISASUGRA and other
currently available codes for generating SUSY mass spectra are
fully reliable at such large values of $\tanb$, especially in the regime of
$\mhalf,\mzero\gsim1\tev$.  First, two-loop corrections to the
effective potential would need to be added to ensure a further
reduction of the scale dependence.  Second, now the Yukawa couplings
of the bottom and the tau become comparable with that of the
top and the contribution to the effective potential from the sbottom
and stau loops becomes of the same order as that from the stops. This
introduces at least two more mass scales and the choice of the right
scale for minimizing the Higgs potential becomes much more complicated.
The variation in the shape and position of the 
region of no-EWSB and the heavy Higgs masses with
$\mt$ and $\mb$ that we discussed above is, in our opinion, probably
just a reflection of the above problems.

One may conclude that, given the technical difficulty involved, the
available codes for computing especially $\mu$ and the heavy Higgs
masses are not yet fully applicable to such large values of $\tanb$,
especially for $\mhalf,\mzero\gsim1\tev$ and we applaud the ongoing
efforts to ameliorate the situation~\cite{adjouadietalpc}.

%%%

On the other hand, the case of $\tanb\sim50$ is an intriguing one
as it corresponds to the infrared quasi-fixed point solution of the
top Yukawa coupling for the Yukawa RGEs when the approximate
unification with $\yb$ and $\ytau$ is further
assumed~\cite{bbop93,fmk93,cpw93}. Additionally, in $SO(10)$--based
models various texture models for mass matrices of leptons and quarks
invariably require large $\tanb\sim50$~\cite{dhr92,rabyrecent}.
Collider phenomenology can also be distinctively
different~\cite{bcdpt97} with the usual multilepton signatures for
SUSY signals now being much reduced, although new signals may appear
involving $\tau$-leptons and $b$-quarks in the final state.

Finally, in Table~\ref{tableone} we present several representative
cases consistent with all experimental and cosmological bounds. All
the mass parameters are given in $\gev$. The Table should be helpful
in a more detailed comparison of our results with other groups.

\TABULAR[h]
% \begin{table}[h] % [p!]
% \centering
% \renewcommand{\arraystretch}{0.95}
% \begin{tabular}
{|c||r|r|r|r|r|r|r|}{
\hline
Case           & RRN1 &RRN2 & RRN3 & RRN4 & RRN5 & RRN6 & RRN7 \\
\hline
$\tanb$         & 10  & 10  & 40  & 45   & 50 & 50   & 50 \\
$\mhalf$        & 300 & 500 & 250 & 600  & 700& 280  & 1000\\
$\mzero$        & 75  &1900 &1225 & 350  & 600& 1250 & 1000\\
\hline
$\mtpole$       & 175 &175  &175  & 175  & 175& 175  &175 \\
$\mbmbsmmsbar$  & 4.25&4.25 &4.25 &4.25  &4.25& 4.25 &4.25 \\
\hline\hline
$\at$         &-556.3 &-886.8&-453.8 &-1028 &-1184 &-495.3 &-1644  \\
$\abot$       &-803.5 &-1245 &-573.5 &-1274 &-1412 &-579.2 &-1937  \\
$\atau$       &-186.0 &-301.5&-74.2  &-166.5 &-143.0 &-38.4 &-219.6 \\
\hline
$\mhl$~(ISA)    & 114.6 &118.0& 115.5&120.0 &120.7 &115.8 & 122.3\\
$\mhl$~(FHF)    & 112.8 &115.1& 113.1&118.3 &119.0 &113.5 & 120.5\\
$\mha$          & 443.6 &1920 & 833.6&601.9 &652.5 &537   & 917.5\\
$\mu(\qstop)$   & 397.3 &299.4& 166.7&701.5 &785.2 &209.5 & 1038\\
\hline
$\mchi$         & 117.5 &196.6& 87.8 &247.9 &291.7 &105.3 & 424.6\\
$\mcharone$     & 215.5 &272.7&129.6 &461.1 &542.4 &163.7 & 784.6\\
$\mgluino$      & 706.3 &1224 &659.5 &1333  &1558  &727.9 & 2176\\
\hline
$\msell$        & 221.0 &1919&1231   &515.9 &757.2 &1258 & 1193\\
$\mselr$        & 140.9 &1905&1227   &394.8 &653.2 &1253 & 1064\\
% $\msnue$        &     &    &    &  &  &    & \\
$\mstauone$     & 132.2 &1889&1035   &254.4 &429.6 &911.0 & 734.1\\
$\mstautwo$     & 224.4 &1911&1140   &503.1 &695.6 &1104  & 1075\\
% $\msnutau$      &  & &  &  &  &    &\\
$\msupl$        & 630.0 &2085&1294 &1219 &1477 &1341 & 2118\\
$\msupr$        & 608.9 &2075&1293 &1180 &1434 &1337 & 2049\\
$\msdl$         & 634.9 &2087&1296 &1221 &1479 &1343 & 2120\\
$\msdr$         & 607.8 &2074&1293 &1175 &1428 &1337 & 2041\\
$\mstopone$     & 468.3 &1398&835.8&948.0&1139 &870.0 & 1634\\
$\mstoptwo$     & 651.9 &1780&1021 &1137 &1335 &1020 & 1869\\
$\msbotone$     & 580.3 &1780&1003 &1049 &1255 &996.5 &1806 \\
$\msbottwo$     & 609.6 &2059&1135 &1123 &1331 &1093  &1877 \\
\hline
$\br( B\ra X_s \gamma ) /10^{-4}$ 
                & 3.4   & 3.6& 2.8 & 2.6& 2.8& 2.6 & 3.0\\
\hline
$\amu/10^{-10}$    
                & 20.5  &1.1 & 12.0& 21.8 &15.0 & 14.4& 7.0\\
\hline
$\abundchi$     & 0.17  &0.20& 0.16& 0.14 &0.13 & 0.17& 0.16\\

\hline
\hline
}
% \end{tabular}
% \caption{ %[]
{\label{tableone} 
Masses and other variables for several representative choices of
parameters. Quantities with mass dimension are given in $\gev$. 
The mass of the lightest scalar Higgs is computed using
ISASUGRA (ISA) and the FeynHiggsFast (FHF) routines. 
}
% \end{table}

Our results show a sizeable difference at large $\tanb\gg10$ when
compared with some other recent analyses~\cite{fmw00,efgos01,eno01}.
First, in Ref.~\cite{fmw00} a similar procedure for computing Higgs
and SUSY mass spectra was used as in our modified version of 
ISASUGRA, which was based on
Ref.~\cite{pbmz97:ref}. In Ref.~\cite{fmw00} somewhat different quark
input masses were used ($\mtpole=174\gev$ and $\mbpole=4.9\gev$) but,
even for that choice, we find that our  version of 
ISASUGRA still produced significantly lower
values for $\mha$ (by some $100\gev$) at large $\tanb\sim50$ than in
Ref.~\cite{fmw00}, which we find somewhat puzzling. 
(This affects the position of the $\ha$-resonance and therefore also
the relic density contours, even though in Ref.~\cite{fmw00} the
latter was computed only in an approximate
way using the package NEUTDRIVER.)
As
regards Refs.~\cite{efgos01,eno01}, we find a sizeable discrepancy
especially in the mass of the heavy Higgs bosons (especially the
pseudoscalar) and therefore also in the position of the $A$-resonance.
We believe that our computation of $\abundchi$, especially near the
$A$-resonance, is somewhat more precise due to our exact treatment of
the $\chi\chi$-annihilation. However, we are convinced that the main
difference comes from the fact that in Refs.~\cite{efgos01,eno01} the
effective Higgs potential was minimized at the scale
$\qzero=\mz$~\cite{battaglia01} 
while in ISASUGRA this is done at $\qstop=\sqrt{\mstopl \mstopr}$ as
discussed in Sect.~\ref{procedure:sec}.
This leads to large differences in the values of $\mu$ and $\bmu$
especially at larger $\mzero$ and $\mhalf$ which were also discussed
in Ref.~\cite{battaglia01}. We have checked
numerically that choosing the minimization scale at $\mz$ typically
gives significantly smaller mass of the pseudoscalar for a given
choice of other parameters. This has a strong effect on the location
of its resonance in the ($\mhalf,\mzero$)--plane and therefore on the
neutralino relic abundance. It also strongly affects the position of
the region of no-EWSB.

%%%%%%%%%%%%%%%%%%%%%%%%%%%%%%%%%%%%%%%%%%%%%%%%%%%%%%%%%%%%%%%%%%%%%%%%%%
%
\section{Summary}\label{summary:sec}

In summary, the interplay of the different experimental
constraints and the cosmologically expected~(\ref{newachi:eq}) and
allowed~(\ref{newachiubound:eq}) regions shows a significant
dependence on $\tanb$. Generally, the region excluded by $\br( B\ra
X_s \gamma )$ grows with $\tanb$ and so does the $2\sigma$ region
favored by the recent measurement of the anomalous magnetic moment of
the muon. The lightest Higgs mass limit $\mhl>113\gev$ gives a lower
bound on $\mhalf$, which however becomes weaker as $\mzero$ and/or $\mt$
grow. Additionally, there still remains some theoretical 
uncertainty in computing $\mhl$, of the order of $2$--$3\gev$. As can
be seen by comparing the $\mhl$ contours of 113 and $115\gev$, this
affects the range of excluded $\mhalf$ by several tens of $\gev$ at
larger $\mt$ and even by nearly hundred~$\gev$ at smaller $\mt$.

At $\tanb\lsim40$, the experimental bounds exclude large parts of the
generic `low-mass' region of both $\mhalf$ and $\mzero$ (below a few
hundred $\gev$~\cite{kkrw94:ref}) of the expected (green)
range of the neutralino relic abundance ($0.1\lsim\abundchi\lsim0.2$)
which itself is strongly
squeezed by the conservative upper bound $\abundchi\lsim0.3$. What remains
are the two narrow bands of expected $\abundchi$ along the wedges of
no-EWSB and $\stauone$-LSP of which the second is (slightly) more
favored by the measurement of $\amu$.

The picture changes dramatically as $\tanb$ approaches $50$.  The
cosmologically expected regions expand significantly both in size and
towards larger $\mhalf$ and $\mzero$. They also become further affected
by the appearance of the resonance due to the pseudoscalar, whose mass
decreases significantly at very large $\tanb\gsim45$, and especially for 
smaller values of the ratio
of the top to bottom masses. Overall, it is clear that, in contrast to
smaller values of $\tanb$, at such large $\tanb$ cosmological
constraints point towards significantly larger gluino and/or sfermion
masses in the $\tev$ range.

NOTE ADDED: After our analysis was completed, we became aware of 
Ref.~\cite{bk01} where the impact on the CMSSM of the recent
highly restrictive determination of $\abundcdm\simeq
0.13\pm0.01$~\cite{netterfieldetal01:ref} was discussed, along with
other constraints, except for $b\ra s\gamma$, for the choice
$\mtpole=175\gev$ and $\mbmbsmmsbar=4.3\gev$~\cite{ckaopc}. 
We find a reasonable
agreement in the overlapping results, in particular in the location of
the resonance of the pseudoscalar $\ha$. Similarly, we find a reasonable
agreement with Refs.~\cite{ls01,ddk01} which were completed after
our paper.
%%%%%%%%%%%%%%%%%%%%%%%%%%%%%%%%%%%%%%%%%%%%%%%%%%%%%%%%%%%%%%%%%%%%%
%
\bigskip

\acknowledgments
LR and RRdA would like to thank P.~Gambino and G.~Giudice for
beneficial discussions regarding large $\tanb$ corrections to $b\ra
s\gamma$. LR further acknowledges conversations with B.~Allanach,
J.~Ellis, W.~Hollik, K.~Matchev, P.~Nath, K.~Olive, F.~Paige,
M.~Quiros, S.~Raby, D.~Tovey, G.~Weiglein as well as the combined
`unnamed physicist' for providing useful comments on various aspects
of the analysis. 
%Our special thanks go to A.~Roszkowska for her 
%invaluable help in preparing the figures.
%%%%%%%%%%%%%%%%%%%%%%%%%%%%%%%%%%%%%%%%%%%%%%%%%%%%%%%%%%%%%%%%%%%%%%%%%%
%
%%%%%%%%%%%%%%%%%%%%%%%%%%%%%%%%%%%%%%%%%%%%%%%%%%%%%%%%%%%%%%%%%%%%%%%%%%

%%%%%%%%%%%%%%%%%%%%%%%%%%%%%%%%%%%%%%%%%%%%%%%%%%%%%%%%%%%%%%%%%%%%%%%%%
%
%

\begin{thebibliography}{99}

\bibitem{jkg96:ref} 
For a review see G. Jungman, M. Kamionkowski and K. Griest,
\prep{267}{195}{1996}.  

\bibitem{goldberg93}
H. Goldberg, \prl{50}{1419}{1983}.

\bibitem{ehnos}
J. Ellis, J.S.~Hagelin, D.V.~Nanopoulos, K.A.~Olive and M.~Srednicki, 
\npb{238}{453}{1984}.

\bibitem{efgos01}
J.~Ellis, T.~Falk, G.~Ganis, K.A.~Olive and M.~Srednicki, 
\plb{510}{236}{2001}.

\bibitem{eno01}
J.~Ellis, D.V.~Nanopoulos and K.A.~Olive, \plb{508}{65}{2001}.

\bibitem{ads0102}
R.~Arnowitt, B.~Dutta and Y.~Santoso, \npb{606}{59}{2001};
R.~Arnowitt, B.~Dutta, B.~Hu and Y.~Santoso, \plb{505}{177}{2001}.

\bibitem{bk01}
V.~Barger and C.~Kao, \hepph{0106189}. 

\bibitem{turnerjune01}
M.S.~Turner, \astroph{0106035}.

\bibitem{primack00}
J.R.~Primack, 
%Nucl.\ Phys.\ Proc.\ Suppl.\ {\bf 87} (2000) 13,
\astroph{0007187}.

\bibitem{krauss01}
L.~Krauss, \astroph{0106149}.

\bibitem{prykeetal01:ref}
C.~Pryke, \etal, The DASI Collaboration, \astroph{0104490}.

\bibitem{netterfieldetal01:ref}
C.B.~Netterfield, \etal, The BOOMERanG Collaboration, \astroph{0011378}.

\bibitem{rosat01}
S.~Borgani, \etal, The ROSAT Collaboration, \astroph{0106428}.

\bibitem{nillesrev}
For a review see, for instance, H.P. Nilles, \prep{110}{1}{1984}. 

\bibitem{kkrw94:ref}
G.L. Kane, C. Kolda, L. Roszkowski, and J.D. Wells,
\prd{49}{6173}{1994}. 

\bibitem{tevatrontworep00} 
V. Barger, C.E.M. Wagner, \etal, Report of the SUGRA Working Group for
Run II of the Tevatron, \hepph{0003154}.

\bibitem{na92}
P.~Nath and R.~Arnowitt, \prl{70}{3696}{1993}.

\bibitem{rr93} 
R.G. Roberts and L. Roszkowski, \plb{309}{329}{1993}.

\bibitem{baerbrhlik96} 
H.~Baer and M.~Brhlik, \prd{53}{597}{1996};
H.~Baer, M.~Brhlik, M.A.~Diaz, J.~Ferrandis, P.~Mercadante, P.~Quintana
and X.~Tata, \prd{63}{015007}{2001}.

\bibitem{fmm99}
J.~L.~Feng, K.~T.~Matchev and T.~Moroi, \prl{84}{2322}{2000},
\hepph{9908309} and \prd{61}{075005}{2000}, \hepph{9909334}. 

\bibitem{lephiggsclaim00}
R.~Barate \etal (ALEPH), \plb{495}{1}{2000}, \hepex{0011045};\\
P.~Abreu \etal (DELPHI), \plb{499}{23}{2001}, \hepex{0102036};\\
M.~Acciarri \etal, (L3), \plb{495}{18}{2000}, \hepex{0011043};\\
G.~Abbiendi \etal, (OPAL), \plb{499}{38}{2001},
\hepex{0101014}.

\bibitem{cqw} 
M.~Carena, J.R.~Espinosa, M.~Quiros and C.E.M.~Wagner,
\plb{355}{209}{1995}, \hepph{9504316}; M.~Carena, 
M.~Quiros and C.E.M.~Wagner, \npb{461}{407}{1996}, \hepph{9508343}.

\bibitem{espinosa} 
J.R.~Espinosa and R.-J.~Zhang, \hepph{9912236} and \hepph{0003246}.

\bibitem{chhhww00}
M.~Carena, \etal, 
%H.E.~Haber, S.~Heinemeyer, W. Hollik, C.E.M. Wagner, G. Weiglein, 
\npb{580}{29}{2000}, \hepph{0001002}.

\bibitem{feynhiggsfast:ref}
S.~Heinemeyer, W.~Hollik and G.~Weiglein, \hepph{0002213}.

\bibitem{misiakmoriond01:ref} 
M.~Misiak, talk at the XXXVIth Rencontres de
Moriond, Les Arcs, March 2001, \hepph{0105312}.

\bibitem{cleo0108} 
S.~Chen, \etal, The CLEO Collaboration, \hepex{0108032}.

\bibitem{hrs:ref}
L.J.~Hall, R.~Rattazzi and U.~Sarid, \prd{50}{7048}{1994}, \hepph{9306309}.

\bibitem{bb98} 
H.~Baer and M.~Brhlik, \prd{58}{015007}{1998}, \hepph{9712305}.

\bibitem{br99} 
T.~Blazek and S.~Raby, \prd{59}{095002}{1999}, \hepph{9712257}.

\bibitem{bhgk00} 
W.~de~Boer, \etal, 
%M. Huber, A.V. Gladyshev, D.I. Kazakov, 
\hepph{0007078} and \hepph{0102163}.

\bibitem{cgnw00:ref}
M.~Carena, D.~Garcia, U.~Nierste and C.E.M.~Wagner, \plb{499}{141}{2001}.

\bibitem{dgg00:ref}  G.~Degrassi, P.~Gambino and G.F.~Giudice, 
\jhep{0012}{009}{2000}.

\bibitem{gm01}
P.~Gambino and M.~Misiak, \hepph{0104034}.

\bibitem{gambinopc} P.~Gambino, private communication.

\bibitem{e821jan01:ref}
H.~N.~Brown, \etal, The Muon ($g-2$) Collaboration, hep-ex/0102017. 

\bibitem{czarneckimoriond01:ref}
A.~Czarnecki, talk at the XXXVIth Rencontres de
Moriond, Les Arcs, March 2001.

\bibitem{narison01:ref}
See Ref.~\protect{\cite{czarneckimoriond01:ref}} and, for example,
S.~Narison, \hepph{0103199}.

\bibitem{isasugra:ref}
H.~Baer, F.E.~Paige, S.D.~Protopopescu and X.~Tata, \hepph{0001086}.

\bibitem{pdb01:ref}
D.E.~Groom, \etal, \epjc{15}{1}{2000}, {\tt http://pdg.lbl.gov/}.

\bibitem{sachrajda00:ref}
G.~Martinelli and C.T.~Sachrajda, \npb{559}{429}{1999}; 
C.T.~Sachrajda, hep-lat/0101003. 

\bibitem{pbmz97:ref}
D.M.~Pierce, J.A.~Bagger, K.~Matchev and R.~Zhang, \npb{491}{3}{1997}.

\bibitem{klns00}
A.~Katsikatsou, A.B.~Lahanas, D.V.~Nanopoulos and V.C.~Spanos,
\plb{501}{2001}{69}.

\bibitem{grz:ref}
G.~Gamberini, G.~Ridolfi and F.~Zwirner, \npb{331}{331}{1990}. 

\bibitem{decarloscasas93:ref}
B.~de Carlos and J.A.~Casas, \plb{309}{320}{1993}.

\bibitem{nrr2}
T.~Nihei, L.~Roszkowski and R.~Ruiz de Austri, in preparation.

\bibitem{gg91}
P.~Gondolo and G.~Gelmini, \npb{360}{145}{1991}.

\bibitem{darksusy00}
P.~Gondolo {\it et. al.}, \astroph{0012234}. 

\bibitem{dn93}
M.~Drees and M.~Nojiri, \prd{47}{376}{1993}.

\bibitem{nrr1}
T.~Nihei, L.~Roszkowski and R.~Ruiz de Austri, \jhep{0105}{063}{2001}.

\bibitem{eg97}
J.~Edsjo and P.~Gondolo, \prd{56}{1879}{1997}.

\bibitem{efos:coann}
J.~Ellis, T.~Falk and K.A.~Olive, \plb{444}{367}{1998};  
J.~Ellis, T.~Falk, K.A.~Olive and M.~Srednicki, \app{13}{181}{2000}.

\bibitem{bkao98}
V.~Barger and C.~Kao, \prd{57}{3131}{1998}.

\bibitem{chatnath96}
U.~Chattopadhyay and P.~Nath, \prd{53}{1648}{1996}, \hepph{9507386}.

\bibitem{lns00}
A.B.~Lahanas, D.V.~Nanopoulos and V.C.~Spanos, \hepph{0009065}.

\bibitem{adjouadietalpc} B.C.~Allanach, \hepph{0104145}; A.~Djouadi and
J.-L.~Kneur, A.~Dedes, F.~Paige, private communication. 

\bibitem{bbop93} V.~Barger, M.S.~Berger, P.~Ohmann and
R.J.N.~Phillips, \plb{314}{351}{1993}.

\bibitem{fmk93} C.D.~Froggatt, R.G.~Moorhouse and I.G.~Knowles, 
\plb{298}{356}{1993}.

\bibitem{cpw93} M.~Carena, S.~Pokorski and C.E.M.~Wagner, 
\npb{406}{59}{1993}.

\bibitem{dhr92} S.~Dimopoulos, L.J.~Hall and S.~Raby,
\prl{68}{1984}{1992}; \prd{45}{4192}{1992}.

\bibitem{rabyrecent} 
T.~Blazek, S.~Raby and K.~Tobe, \prd{60}{113001}{1999} and
\prd{62}{055001}{2000}.

\bibitem{bcdpt97} H.~Baer, 
C.~Chen, M.~Drees, F.~Paige and X.~Tata, \prl{79}{986}{1997}.

\bibitem{fmw00}
J.L.~Feng, K.~T.~Matchev and F.~Wilczek, \plb{482}{388}{2000} and
\prd{63}{045024}{2001}.

\bibitem{battaglia01}
M.~Battaglia, \etal, hep-ph/0106204.

\bibitem{ckaopc}
C.~Kao, private communication.

\bibitem{ls01}
A.B.~Lahanas and V.C.~Spanos, \hepph{0107151}.

\bibitem{ddk01}
A.~Djouadi, M.~Drees and J.L.~Kneur, \hepph{0107316}.


\end{thebibliography}
\end{document}